\documentclass[aps,preprint,floats,epsf,epsfig,nofootinbib,letter]{revtex4}

\usepackage{graphicx}
\usepackage{dcolumn}
\usepackage{bm}

\begin{document}
\def\be{\begin{eqnarray}}
\def\en{\end{eqnarray}}
\def\la{\langle}
\def\ra{\rangle}
\def\non{\nonumber}
\def\B{{\cal B}}
\def\ov{\overline}
\def\up{\uparrow}
\def\dw{\downarrow}
\def\vp{\varepsilon}
\def\CP{{\it CP}~}
\def\pr{{Phys. Rev.}~}
\def\prl{{ Phys. Rev. Lett.}~}
\def\pl{{ Phys. Lett.}~}
\def\np{{ Nucl. Phys.}~}
\def\zp{{ Z. Phys.}~}
\newcommand{\acp}{\ensuremath{A_{CP}}}
\def\lsim{ {\ \lower-1.2pt\vbox{\hbox{\rlap{$<$}\lower5pt\vbox{\hbox{$\sim$}
}}}\ } }
\def\gsim{ {\ \lower-1.2pt\vbox{\hbox{\rlap{$>$}\lower5pt\vbox{\hbox{$\sim$}
}}}\ } }

\font\el=cmbx10 scaled \magstep2{\obeylines\hfill June, 2015}

\vskip 1.5 cm

\centerline{\large\bf Revisiting Scalar Glueballs}

\bigskip
\bigskip
\centerline{\bf Hai-Yang Cheng$^1$, Chun-Khiang Chua$^2$
and Keh-Fei Liu$^3$}
\medskip
\centerline{$^1$ Institute of Physics, Academia Sinica}
\centerline{Taipei, Taiwan 115, Republic of China}
\medskip
\centerline{$^2$ Department of Physics and Center for High Energy Physics}
\centerline{Chung Yuan Christian University}
\centerline{Chung-Li, Taiwan 320, Republic of China}

\medskip
\centerline{$^3$ Department of Physics and Astronomy, University
of Kentucky} \centerline{Lexington, KY 40506}
\medskip
\bigskip
\bigskip
\centerline{\bf Abstract}
\bigskip
\small

It is commonly believed that the lowest-lying scalar glueball lies somewhere in the isosinglet scalar mesons $f_0(1370), f_0(1500)$ and  $f_0(1710)$ denoted generically by $f_{0}$. In this work we consider lattice calculations and experimental data to infer the glue and $q\bar q$ components of $f_0$. These include the calculations of the scalar glueball masses in quenched and unquenched lattice QCD, measurements of the radiative decays $J/\psi\to\gamma f_{0}$, the ratio of $f_{0}$ decays to $\pi\pi$, $K\ov K$ and $\eta\eta$, the ratio of $J/\psi$ decays to $f_0(1710)\omega$ and $f_0(1710)\phi$, the $f_0$ contributions to $B_s\to J/\psi\pi^+\pi^-$, and the near mass degeneracy of $a_0(1450)$ and $K_0^*(1430)$. All analyses suggest the prominent glueball nature of $f_0(1710)$ and the flavor octet structure of $f_0(1500)$.

\pagebreak

\section{Introduction}
The existence of glueballs is an archetypal prediction of QCD as a confining theory.
It is generally believed that the lowest-lying scalar glueball lies somewhere in the isosinglet scalar mesons with masses above 1 GeV. To see this we first give a short review on scalar mesons (see e.g. \cite{Close:2002,Klempt,Amsler,Liu:2007}).
Many scalar mesons with masses lower than 2 GeV have been observed and they can be classified into two nonets: one nonet with mass below or close to 1 GeV, such as $f_0(500)$ (or $\sigma$), $K_0^*(800)$ (or $\kappa$), $f_0(980)$ and $a_0(980)$  and the other nonet with mass above 1 GeV such as $K_0^*(1430)$, $a_0(1450)$ and two isosinglet scalar mesons. Of course, the two nonets cannot be both low-lying $^3P_0$ $q\bar q$ states simultaneously. If the light scalar nonet is identified with the P-wave $q\bar q$ states, one will encounter two major difficulties: First, why are $a_0(980)$ and $f_0(980)$ degenerate in their masses? In the two quark model, the latter is dominated by the $s\bar s$ component, whereas the former cannot have the $s\bar s$ content since it is an $I=1$ state. Second, why are $f_0(500)$ and $K_0^*(800)$ so broad compared to the narrow widths of $a_0(980)$ and $f_0(980)$ even though they are all in the same nonet? These difficulties with mass degeneracy and the hierarchy of widths can be easily overcome in the tetraquark model \cite{Jaffe}. Therefore, this suggests that the heavy scalar nonet is composed of P-wave $q\bar q$ states, while the light nonet is made of S-wave tetraquark states.

Final-state interactions of $\pi\pi$, $K\ov K$, $\cdots$ etc., are known to be very important in the region below 2 GeV. Such interactions can be described in unitarized chiral perturbation theory (ChPT) or unitarized quark models with coupled channels. It follows that the light scalar mesons  $\sigma$, $\kappa$, $f_0(980)$ and $a_0(980)$ can be dynamically generated through pseudoscalar meson-pseudoscalar meson scattering within the framework of unitarized ChPT valid up to 1.2 GeV  (see \cite{Pelaez} and references therein).
\footnote{A coupled channel study of the meson-meson $S$ wave in terms of 13 coupled channels in \cite{Albaladejo:2008qa} indicates that all the resonances with masses below 2 GeV and $I=0$ and 1/2 can be dynamically generated.}
This implies that these light scalars may have non-negligible contents of hadronic molecules.
The dynamically generated bound state or resonance is characterized by a strong coupling to the coupled channel. For example, both $f_0(980)$ and $a_0(980)$ have been advocated to be $K\ov K$ molecular states \cite{Weinstein,Locher}, while $f_0(500)$ a $\pi\pi$ resonance. By the same token, it has been shown that $f_0(1370)$ and $f_0(1710)$ can be dynamically generated from the $\rho\rho$ interaction in a hidden gauge unitary approach \cite{Molina,Geng}. That is, they have $\rho\rho$ molecular components in addition to the $q\bar q$ content.

Although the light scalar nonet is composed of tetraquark and/or molecular states, it is allowed to have a small amount of the $q\bar q$ component for several reasons: (i) A mixing of the heavy $q\bar q$ scalar nonet with the light nonet will enable us to understand the near degeneracy of $a_0(1450)$ and $K_0^*(1430)$ \cite{Black}. (ii) The large $N_c$ dependence of unitarized two-loop ChPT partial waves for the description of pion-pion scattering suggests a subdominant $q\bar q$ component of the $f_0(500)$ possibly originates around 1 GeV \cite{Pelaez:2006}. (iii) If $f_0(980)$ is a loosely bound state of $K\ov K$, it will be hard to understand its prompt production in $B$ decays. This will require an $s\bar s$ core component in $f_0(980)$. Likewise, the heavy scalar nonet dominated by $q\bar q$ can have molecular and tetraquark components.

In principle, two-quark, four-quark and molecular components of light and heavy scalar mesons can be studied in lattice QCD with the corresponding interpolating fields. So far, the lattice calculation with all the interpolating fields available at the same time is not yet practical (for a review of previous works for light scalar mesons in full lattice QCD, see \cite{Wakayama}). For heavier scalar mesons, the masses of $a_0(1450)$ and $K_0^*(1430)$ have been calculated using the two-quark interpolation field ${\it \ov\Psi\!\Psi}$ \cite{Mathur:2006}. The chirally extrapolated masses $1.42\pm0.13$ GeV for $a_0$ and $1.41\pm0.12$ GeV for $K_0^*$ suggest that the mesons $a_0(1450)$ and $K_0^*(1430)$ are predominantly $q\bar q$ states.

Taking the lattice result as a cue, we shall assume in this work that the scalar meson nonet above 1 GeV is primarily a  $q\bar q$ state in nature. To the lowest order approximation we will not consider the possible tetraquark and molecular contributions. Experimentally, there exist three isosinglet scalars $f_0(1710)$, $f_0(1500)$, $f_0(1370)$ above 1 GeV. They cannot be all accommodated in the $q\bar q$ nonet picture. One of them could be primarily a scalar glueball. It has been suggested that
$f_0(1500)$ is predominately a scalar glueball in \cite{Close1}. Lattice calculations indicate that the mass of the low-lying scalar glueball lies in the range of $1.5- 1.8$ GeV (see Table \ref{tab:Gmass} below). This suggests that $f_0(1370)$ does not have a sizable glue content.   Among the two remaining isoscalar mesons, $f_0(1500)$ and $f_0(1710)$, it has been quite controversial as to which of the two is the dominant scalar glueball.  Since the glueball is hidden somewhere in the quark sector, this is the main reason why the glueball is so elusive.

It is worthy mentioning that the very existence of $f_0(1370)$ has long been considered to be questionable (see e.g. \cite{Klempt} and \cite{Ochs} for detailed discussions). Its mass and width are quoted by PDG (Particle Data Group) \cite{PDG} to be $1200-1500$ MeV and $200-500$ MeV, respectively.
It appears that the decays into two pion isobar can be described by the two poles $f_0(1500)$ and $f_0(1710)$, while four pion isobar can be also described by the two poles $f_0(1370)$ and $f_0(1710)$. However, there is no any single publication showing the need of three states simultaneously. Hence,
the hypothesis of three distinct poles  $f_0(1370)$, $f_0(1500)$ and $f_0(1710)$ is not a general consensus and there is (probably) not a single experiment favoring this hypothesis.

In spite of the controversies on the identification of the scalar glueball, the 2006 version of PDG  \cite{PDG2006} attempted to conclude the status as ``Experimental evidence is mounting that $f_0(1500)$ has considerable affinity for glue and that the $f_0(1370)$ and $f_0(1710)$ have large $u\bar u+d\bar d$ and $s\bar s$ components, respectively". This has been toned down to ``The $f_0(1500)$ or, alternatively, the $f_0(1710)$ have been proposed as candidates for the scalar glueball" in the latest version of PDG.

Using the CLEO data, Dobbs {\it et al.} \cite{Dobbs:2015dwa} have recently analyzed the radiative decays of $J/\psi$ and $\psi(2S)$ into $\pi\pi$, $K\ov K$ and $\eta\eta$. They have determined the product branching fractions for the radiative decays of $J/\psi$ and $\psi(2S)$ to scalar resonances such as $f_0(1370),f_0(1500), f_0(1710)$ and found (see also Table \ref{tab:Jrad})
\be \label{eq:R1710}
R(f_0(1710))\equiv {\Gamma(f_0(1710)\to\pi\pi)\over \Gamma(f_0(1710)\to K\ov K)}=0.31\pm0.05\,.
\en
For a pure, unmixed glueball, its decays to pseudoscalar pairs are expected to be flavor blind. Hence, decays to $\pi\pi$, $K\ov K$, $\eta\eta$, $\eta'\eta'$ and $\eta\eta'$ should have branching fractions proportional to $3:4:1:1:0$ apart from the phase space factor. 
\footnote{For a pure, unmixed glueball, the ratio $R(G)$ defined in Eq. (\ref{eq:RG}) below approaches to 3/4 in the SU(3) limit. Taking into account of phase space corrections, we find $R(G)=0.90$ for $M_G=1710$ MeV and 0.98 for $M_G=1500$ MeV.}
Therefore, Dobbs {\it et al.} concluded that $f_0(1710)$ is not a pure scalar glueball. By the same token, the large deviation of the experimental measurement \cite{PDG}
\be \label{eq:R1500}
R(f_0(1500))\equiv {\Gamma(f_0(1500)\to\pi\pi)\over \Gamma(f_0(1500)\to K\ov K)}=4.1\pm0.5
\en
from the value of 3/4 also implies that $f_0(1500)$ cannot be a pure glueball either.

Denoting $N\equiv n\bar n=(u\bar u+d\bar d)/\sqrt{2}$ and $S\equiv s\bar s$, we write
\be \label{eq:f0}
|f_{0i}\ra=\alpha_{i}|N\ra+\beta_{i}|S\ra+\gamma_{i}|G\ra
\en
with $f_{0i}$ being $f_0(1370),f_0(1500),f_0(1710)$, respectively, for $i=1,2,3$. At first sight, it appears that Eq. (\ref{eq:R1710}) implies $\alpha_3< \beta_3$ while Eq. (\ref{eq:R1500}) leads to $\alpha_2> \beta_2$. However, this may be misleading because the $n\bar n$ component contributes to both $\pi\pi$ and $K\ov K$, while $s\bar s$ contributes only to $K\ov K$. Therefore, it is possible to accommodate $R(f_0(1500))$ even with $|\alpha_2|<|\beta_2|$.

The above-mentioned flavor blindness of glueball decays is valid for $J\neq 0$ glueballs. For a scalar glueball, it cannot decay into a quark-antiquark pair in the chiral limit (see Sec. III.C below for discussion). Consequently, a large suppression of the $\pi\pi$ production relative to $K\ov K$ is expected in the spin-0 glueball decay, though it is difficult to quantify the effect of chiral suppression. Therefore, the ratio $R(G)$ to be defined in Eq. (\ref{eq:RG}) below will be naturally small. Comparison of this with Eqs. (\ref{eq:R1710}) and  (\ref{eq:R1500}) suggests that $f_0(1710)$ is likely to have a large glueball component.

In the literature, there exist two different types of models for the mixing between the scalar glueball $|G\ra$ and the scalar quarkonia $|N\ra$ and $|S\ra$ (see \cite{Mathieu,Crede,Ochs} for reviews). In the first type of models,  $f_0(1500)$ is composed primarily of a glueball with large mixing with $q\bar q$ states, $f_0(1710)$ is predominately a $s\bar s$
state and $f_0(1370)$ is dominated by the $n\bar n$ content. In contrast, in the second type of models, $f_0(1710)$ is primarily a glueball state and $f_0(1500)$ is dominated by the $s\bar s$ component, while $f_0(1370)$ is still governed by the $n\bar n$.

In our previous work \cite{CCL}, we have employed two simple and robust
results as inputs for the mass matrix which is essentially the
starting point for the mixing model between scalar quarkonia and the
glueball. We have shown that $f_0(1710)$ is composed primarily of a scalar glueball. In this work, we shall point out that new results from the unquenched lattice QCD calculation of the glueball spectrum, new measurements of radiative decays of $J/\psi$, a new lattice calculation of $J/\psi\to\gamma G$ and new experimental results on the scalar meson contribution to $B_s\to J/\psi\pi^+\pi^-$ all support the prominent glueball nature of $f_0(1710)$.

This work is organized as follows. In Sec. II, we first outline the
general expected features of a pure glueball and then discuss two different types of models for the mixing between the glueball and quarkoina states. We proceed to discuss various signals for the existence of a scalar glueball, such as the lattice calculations of the glueball spectrum, the radiative decays of $J/\psi$ to isosinglet scalar mesons, $\cdots$, etc. In the vicinity of $f_0(1710)$ there exist several possible other $0^{++}$ states. Their mixing effects are briefly discussed in Sec. IV.
Discussion and conclusions are presented in Sec. V.

\section{Model for Scalar glueball-quarkonia mixing}

A pure glueball state is expected to exhibit the following signatures (see e.g. \cite{Amsler:2006})

\begin{enumerate}
\item It is produced copiously in the glue-rich environment such as radiative $J/\psi$ decays $J/\psi\to\gamma gg$ (or $Q\bar Q\to \gamma gg$ \cite{Nussinov})
      as the glueball couples strongly to the color-singlet digluon.
\item It is suppressed in $\gamma\gamma$ reactions.
\item  Its width is commonly believed to be narrow, say, of order 100 MeV,  as inferred from the large-$N_c$ argument that the
    glueball decay width scales as $1/N_c^2$, while the width of the $q\bar q$ state is $\propto 1/N_c$. Hence, the very broad $f_0(500)$ does not appear to be a good scalar glueball candidate.
\item  The decay amplitude for $J\neq 0$ glueballs is flavor symmetric, namely, its coupling is flavor independent \cite{Close1}.
    A scalar glueball cannot decay into  a massless quark pair or a photon pair to leading order. Hence, its decay amplitude is subject to chiral suppression (see Sec. III.C below for detailed discussions and references).  However, this feature does not hold for pseudoscalar glueballs owing to the axial anomaly \cite{Frere}.
    Consequently, the scalar glueball decay to mesons is sensitive to flavor or SU(3) breaking.

\end{enumerate}
The above features provide qualitative criteria for distinguishing glueballs from $q\bar q$ states with the same quantum numbers. The suppression in $\gamma\gamma$ reactions is usually not a good criterion because the quark mixing can be adjusted in such a way that the $q\bar q$ state has a weak or even vanishing coupling to two photons.

A physical glueball state is an admixture of the glueball with the $q\bar q$ state or even the tetraquark state with the same quantum numbers so that a pure glueball is not likely to exist in nature. In the following we shall consider
two different types of models for the mixing of the scalar glueball with the scalar quarkonia:

\vskip 0.4cm
{\bf (i) Model I: $f_0(1500)$ as primarily a scalar glueball}
\vskip 0.2cm
Amsler and  Close \cite{Close1} claimed $f_0(1500)$ discovered at LEAR as an evidence for a scalar glueball because its decay to $\pi\pi,K\ov K,\eta\eta,\eta\eta'$ is not compatible with a simple $q\bar q$ picture. This is best illustrated in the argument given by Amsler \cite{Amsler02}.
Let $|f_0(1500)\ra=\cos\alpha |N\ra-\sin\alpha|S\ra$. The
suppression of the $K\ov K$ production relative to $\pi\pi$  (cf. Eq. (\ref{eq:R1500}))
indicates that $f_0(1500)$ is $n\bar n$ dominated. This is also well established in $pp$ and $p\bar p$ collisions. By contrast, the non-observation of $f_0(1500)$ in $\gamma\gamma$ reactions implies that $f_0(1500)$ is $s\bar s$ dominated. This is because $\Gamma_{\gamma\gamma}\propto (5\cos\alpha-\sqrt{2}\sin\alpha)^2$ (see Eq. (\ref{eq:2gamma}) below), and hence a small rate implies that $\alpha$ is close to $75^\circ$. Obviously, the above two conclusions
are in contradiction. This led Amsler to argue that $f_0(1500)$ is not a $q\bar q$ state but rather something else and suggested that it is primarily a glueball. This can explain why its $\gamma\gamma$ coupling is weak and why it is produced abundantly in $pp$ and $p\bar p$ collisions. However, this interpretation has a difficulty with the large suppression of $K\ov K$ production relative to $\pi\pi$.

A typical result of the mixing
matrices obtained by Amsler, Close and Kirk~\cite{Close1}, Close
and Zhao~\cite{Close2}, He {\it et al.}~\cite{He:2006} and Yuan {\it et al.} \cite{Yuan} is the following
 \be \label{eq:Close}
 \left(\matrix{ |f_0(1370)\ra \cr |f_0(1500)\ra \cr |f_0(1710)\ra \cr}\right)=
\left( \matrix{ -0.91 & -0.07 & 0.40 \cr
                 -0.41 & 0.35 & -0.84 \cr
                0.09 & 0.93 & 0.36 \cr
                  } \right)\left(\matrix{|N\ra \cr
 |S\ra \cr |G\ra \cr}\right),
 \en
taken from \cite{Close2}.
Eq. (\ref{eq:Close}) will be referred as Model I.  A common feature of these analyses is that, before mixing, the
$s\bar{s}$ quarkonium mass $M_S$ is larger than the glueball mass $M_G$
which, in turn, is larger than the
$n\bar n$ quarkonium mass $M_N$, with $M_G$
close to 1500 MeV and $M_S-M_N$ of the order of $200\sim 300$ MeV.
In this model, $f_0(1710)$ is considered mainly as a $s\bar s$
state, while $f_0(1370)$ is dominated by the $n\bar n$ content and
$f_0(1500)$ is composed primarily of a glueball with possible
large mixing with $q\bar q$ states.

\vskip 0.4cm
{\bf (ii) Model II: $f_0(1710)$ as primarily a scalar glueball}
\vskip 0.2cm
Based on the lattice calculations, Lee and Weingarten \cite{Lee}
found that $f_0(1710)$ to be composed mainly of the scalar
glueball, $f_0(1500)$ is dominated by the $s\bar s$ quark content,
and $f_0(1370)$ is mainly governed by the $n\bar n$ component, but
it also has a glueball content of 25\%. Their mixing matrix is
 \be \label{eq:Lee}
 \left(\matrix{ |f_0(1370)\ra \cr |f_0(1500)\ra \cr |f_0(1710)\ra \cr}\right)=
\left( \matrix{ 0.819(89) & 0.290(91) & -0.495(118) \cr
                 -0.399(113) & 0.908(37) & -0.128(52) \cr
                0.413(87) & 0.302(52) & 0.859(54) \cr
                  } \right)\left(\matrix{|N\ra \cr
 |S\ra \cr |G\ra \cr}\right).
 \en
In this scheme, $M_S=1514\pm11$ MeV, $M_N=1470\pm25$ MeV and
$M_G=1622\pm29$ MeV.

To improve this model, it is noted in \cite{CCL} that two crucial facts need to be incorporated as the starting point for the mixing calculation.  First of all, it is known empirically that flavor SU(3) is an approximate symmetry in the scalar meson sector above 1 GeV. The multiplets of the light scalar mesons $K_0^*(1430)$, $a_0(1450)$ and $f_0(1500)$ are nearly degenerate. In the scalar charmed meson sector, $D_{s0}^*(2317)$ and $D_0^*(2400)$
\footnote{In spite of its notation, the mass of $D_0^*(2400)^0$, $2318\pm29$ MeV \cite{PDG}, is almost identical to the mass of $D_{s0}^*(2317)$, $2317.8\pm0.6$ MeV.}
have very similar masses even though the former contains a strange quark. It is most likely that the same phenomenon also holds in the scalar bottom meson sector \cite{Cheng:2014}. This unusual behavior is not understood
as far as we know and it serves as a challenge to the existing
hadronic models, but the degeneracy of $a_0(1450)$ and
$K_0^*(1430)$ is confirmed in the quenched lattice calculation~\cite{Mathur:2006}.
This requires that there not be a $\sim$ 200 MeV difference between the
$s\bar{s}$ state and the $n\bar{n}$ in the diagonal matrix elements in the mixing matrix as have been done in all the previous calculations.
Second, a latest quenched lattice calculation
of the glueball spectrum at the infinite volume and continuum
limits based on much larger and finer lattices have been carried
out~\cite{Chen:2005mg}. The mass of the scalar glueball is calculated to
be $m(0^{++})=1710\pm50\pm 80$ MeV.   This suggests that $M_G$
should be close to 1700 MeV rather than 1550 MeV from the earlier
lattice calculations~\cite{Bali}.

We begin by considering exact SU(3) symmetry as a first
approximation for the mass matrix, namely, $M_S=M_U=M_D=M$ with  $M_{U,D,S}$ being the masses of the scalar quarkonia
$u\bar u$, $d\bar d$ and $s\bar s$, respectively, before mixing.
In this case, two of the mass eigenstates are to be identified
with $a_0(1450)$ and $f_0(1500)$ which are degenerate with the
mass $M$ before mixing. Taking $M$ to be the experimental mass of $1474\pm 19$
MeV of $a_0(1450)$, it is a good approximation for the mass of
$f_0(1500)$ at $1505\pm 6$ MeV \cite{PDG}. Thus, in the limit of
exact SU(3) symmetry, $f_0(1500)$ is an SU(3) isosinglet octet
state $|f_{\rm octet}\ra= {1\over\sqrt{6}}(|u\bar u\ra+|d\bar d\ra-2|s\bar
  s\ra)={1\over\sqrt{3}}(|N\ra-\sqrt{2}|S\ra)$
and is degenerate with $a_0(1450)$. In the absence of glueball-quarkonium mixing, $f_0(1710)$ would be a pure glueball and $f_0(1370)$ a pure SU(3) singlet $|f_{\rm singlet}\ra={1\over\sqrt{3}}(|u\bar u\ra+|d\bar d\ra+|s\bar s\ra)={1\over\sqrt{3}}(\sqrt{2}|N\ra+|S\ra)$ and its mass is shifted down by 3 times the coupling between the $u\bar{u},
d\bar{d}$ and $s\bar{s}$ states which
is $\sim 100$ MeV lower than $M$. When the
glueball-quarkonium mixing is turned on, there will be additional
mixing between the glueball and the SU(3)-singlet $q\bar{q}$ . As a result, the mass
shift of $f_0(1370)$ and $f_0(1710)$ due to this mixing is only of order 10 MeV. Since the SU(3) breaking effect is expected to be
weak, it can be treated perturbatively. The obtained mixing matrix is \footnote{We have updated the fit results in \cite{CCL} by taking into account the experimental uncertainties of the isosinglet scalar meson masses and branching fractions.  The other updated parameters in fit (ii) are $r_a=1.21^{+0.07}_{-0.09}$, $\rho_s=0.12^{+0.02}_{-0.05}$ and $\rho_{ss}=0.60^{+1.24}_{-2.02}$.}
\be \label{eq:CCL}
 \left(\matrix{ |f_0(1370)\ra \cr |f_0(1500)\ra \cr |f_0(1710)\ra \cr}\right)=
\left( \matrix{ 0.78\pm0.02 & 0.52\pm0.03 & -0.36\pm0.01 \cr
                 -0.55\pm0.03 & 0.84\pm0.02 & 0.03\pm0.02 \cr
                0.31\pm0.01 & 0.17\pm0.01 & 0.934\pm0.004 \cr
                  } \right)\left(\matrix{|N\ra \cr
 |S\ra \cr |G\ra \cr}\right)
\en
with $M_N=1474$ MeV, $M_S=1496\pm14$ MeV and $M_G=1674\pm14$ MeV
will be referred as Model II.
It is evident that $f_0(1710)$ is composed primarily of the scalar
glueball, $f_0(1500)$ is close to an SU(3) octet, and $f_0(1370)$
consists of an approximated SU(3) singlet with some glueball
component ($\sim 10\%$). Unlike $f_0(1370)$, the glueball content
of $f_0(1500)$ is very tiny because an SU(3) octet does not mix
with the scalar glueball.

For other glueball-quarkonium mixing models in this category, namely, $f_0(1710)$ is predominantly a glueball, see \cite{Janowski}.

\section{Signal for Scalar glueball and its mixing with quarkonium}
In this section we shall consider
the calculations of the scalar glueball mass in quenched and unquenched lattice QCD, the radiative decay $J/\psi\to\gamma f_0$, the ratio of $f_0$ decays to $\pi\pi$, $K\ov K$ and $\eta\eta$, the ratio of $J/\psi$ decays to $f_0(1710)\omega$ and $f_0(1710)\phi$, the scalar contributions to $B_s\to J/\psi\pi^+\pi^-$, and the near mass degeneracy of $a_0(1450)$ and $K_0^*(1430)$. They will provide clues on the coefficients $\alpha_i,\beta_i$ and $\gamma_i$ in Eq. (\ref{eq:f0}) for isosinglet scalar mesons $f_{0i}$. For example, the radiative decay $J/\psi\to\gamma f_{0i}$ is sensitive to the glue content of $f_{0i}$, while the study of scalar contributions to $B_s\to J/\psi\pi^+\pi^-$ can be used to explore the $s\bar s$ component of $f_{0i}$. For the study of the scalar glueball production in hadronic $B$ decays, see \cite{He:2015}.

\subsection{Masses from lattice calculations}

Lattice calculations of the scalar glueball mass in quenched and unquenched QCD are summarized in Table \ref{tab:Gmass}. Except for the earlier calculation by Bali {\it et al.} \cite{Bali}, the mass of a pure gauge scalar glueball falls in the range of 1650$-$1750 MeV. The latest quenched lattice calculation of the glueball spectroscopy by Chen {\it et al.} \cite{Chen:2005mg} shows that the lightest scalar glueballs has a  mass of order 1710 MeV. The predicted masses in quenched lattice QCD are for pure glueballs in the Yang-Mills gauge theory. The question is what happens to the glueballs in the presence of quark degrees of freedom? Is the QCD glueball heavier or lighter than the one in Yang-Mills theory? In full QCD lattice calculations, glueballs will mix with fermions, so pure glueballs does not exist. The unquenched calculation carried out in \cite{Gregory} gives $1795\pm60$ MeV for the lowest-lying scalar glueball.
\footnote{An earlier full QCD lattice study in \cite{UKQCD} did not give  numerical results on glueball masses except in the last figure of the paper.
In unquenched lattice QCD, the glueball is not the lowest state. There are other mesons below it. This makes it harder to isolate and identify the glueball. Hence, there are not many unquenched calculations.}
It suggests that the unquenching effect is small; the mass of the scalar glueball is not significantly affected by the quark degree of freedom.

It is clear that both quenched and unquenched lattice calculations indicate that $f_0(1710)$ should have a large content of the scalar glueball. In principle, the percentage of the $0^{++}$ glue component in $f_0(1710)$ can be calculated in full lattice QCD by considering the overlap of $f_0(1710)$ with the glue and $q\bar q$ operators. \footnote{Notice that quenched lattice QCD has been used in \cite{Lee} to estimate the mixing between the glue and $q\bar q$ states.}

In the glueball-quarkonia mixing models considered in Sec. II, the parameter $M_G$ is the mass of the scalar glueball in the pure gauge sector. In Model I, $M_G=1464\pm47$ MeV in  fit 1 and $1519\pm41$ MeV in fit 2 \cite{Close2}, while it is of order 1665 MeV in Model II \cite{CCL}. Obviously, the latter lies in the range of quenched lattice results for a pure scalar glueball.

\begin{table}[t]
\caption{Scalar glueball masses (in units of MeV) in quenched (top) and unquenched (bottom) lattice QCD. } \label{tab:Gmass}
\begin{tabular}{ | l | c   |}  \hline
 ~~~Bali {\it et al.} (1993) \cite{Bali} & $1550\pm 50$ \\
 ~~~H. Chen {\it et al.} (1994) \cite{Chen:1994} & $1740\pm71$\\
 ~~~Morningstar, Peardon (1999) \cite{Morningstar} & ~~$1730\pm50\pm80$~~ \\
 ~~~Vaccarino, Weingarten (1999) \cite{Vaccarino} ~~ & $1648\pm 58$   \\
 ~~~Loan {\it et al.} (2005) \cite{Loan} & $1654\pm83$ \\
 ~~~Y. Chen {\it et al.} (2006) \cite{Chen:2005mg} & $1710\pm50\pm80$ \\ \hline
 ~~~Gregory {\it et al.} (2012) \cite{Gregory} & $1795\pm60$ \\ \hline
\end{tabular}
\end{table}

\subsection{Radiative $J/\psi$ decays}
The radiative decay $J/\psi\to \gamma f_0$ is an ideal place to
test the scalar glueball content of $f_0$ since the leading
short-distance mechanism for the inclusive decay $J/\psi\to\gamma+X$ is
$J/\psi\to \gamma+gg$. If $f_0(1710)$ is composed mainly of the
scalar glueball, it should be the most prominent scalar produced
in radiative $J/\psi$ decays. Hence, it is expected that
 \be   \label{1710_1500}
 \Gamma(J/\psi\to \gamma f_0(1710))\gg \Gamma(J/\psi\to \gamma
 f_0(1500)).
 \en

Branching fractions of radiative decays of $J/\psi$ to $f_0(1500)$ and $f_0(1710)$ measured by BES and CLEO are listed in Table \ref{tab:Jrad}.
When summing over various channels in the table, we obtain
\be
\B(J/\psi\to \gamma f_0(1500))> \B(J/\psi\to \gamma f_0(1500)\to \gamma(\pi\pi,\eta\eta))=(1.3\pm0.3)\times 10^{-4},
\en
and
\be
\B(J/\psi\to \gamma f_0(1710))> \B(J/\psi\to \gamma f_0(1710)\to \gamma(\pi\pi,K\ov K,\omega\omega,\eta\eta))=(16.5\pm1.4)\times 10^{-4},
\en
where we have used the average of BES and CLEO measurements whenever both available. It is clear that the lower limit for the radiative decay of $f_0(1710)$ is one order of magnitude larger than $f_0(1500)$. Using the measured branching fractions $\B(f_0(1500)\to\pi\pi)=0.349\pm0.023$ and $\B(f_0(1500)\to\eta\eta)=0.051\pm0.009$ \cite{PDG}, we find
\be \label{eq:gamma1500}
\B(J/\psi\to \gamma f_0(1500))=\cases{(3.13\pm0.73)\times 10^{-4} & from $f_0(1500)\to\pi\pi$, \cr
(3.23\pm2.03)\times 10^{-4} & from $f_0(1500)\to\eta\eta$. \cr}
\en
Likewise, we have
\be \label{eq:gamma1710}
\B(J/\psi\to \gamma f_0(1710))=\cases{(3.27\pm1.88)\times 10^{-3} & from $f_0(1710)\to\pi\pi$, \cr
(2.80\pm0.96)\times 10^{-3} & from $f_0(1710)\to K\ov K$, \cr
}
\en
where the branching fractions  $\B(f_0(1710)\to K\ov K)=0.36\pm0.12$ and $R(f_0(1710))=0.32\pm0.14$ \cite{Albaladejo:2008qa} have been used.
\footnote{For the sake of consistency, we use the results of \cite{Albaladejo:2008qa} for both $\B(f_0(1710)\to K\ov K)$ and $R(f_0(1710))$
obtained from the same data analysis.}
Therefore, we conclude that
\be
 {\Gamma(J/\psi\to \gamma f_0(1710))\over \Gamma(J/\psi\to \gamma
 f_0(1500))}\sim {\cal O}(10) .
\en

The radiative decay of $J/\psi$ to a scalar glueball has been studied by the CLQCD Collaboration within the framework of quenched lattice QCD \cite{CLQCD}. The result is
\be
\B(J/\psi\to \gamma G)=(3.8\pm0.9)\times 10^{-3}.
\en
Comparing this with Eqs. (\ref{eq:gamma1500}) and (\ref{eq:gamma1710}),
it is edvident that $f_0(1710)$ has a larger overlap with the pure glueball than other scalar mesons as expected in Model II.

\begin{table}[t]
\caption{Branching fractions (in units of $10^{-4}$) of radiative decays of $J/\psi$ to $f_0(1500)$ and $f_0(1710)$ measured by BES and CLEO.} \label{tab:Jrad}
\begin{ruledtabular}
\begin{tabular}{|l c c| }
Decay Mode & BES & CLEO \cite{Dobbs:2015dwa} \\ \hline
$J/\psi\to\gamma f_0(1500)\to\gamma \pi\pi$ & $1.01\pm0.32$ \cite{PDG} & $1.21\pm0.29\pm0.24$ \\
$J/\psi\to\gamma f_0(1500)\to\gamma \eta\eta$ & $0.165^{+0.026+0.051}_{-0.031-0.140}$ \cite{BES:etaeta} & $$ \\
\hline
$J/\psi\to\gamma f_0(1710)\to\gamma \pi\pi$ & $4.0\pm1.0$ \cite{PDG} & $3.71\pm0.30\pm0.43$ \\
$J/\psi\to\gamma f_0(1710)\to\gamma K\ov K$ & $8.5^{+1.2}_{-0.9}$ \cite{PDG} & $11.76\pm0.54\pm0.94$ \\
$J/\psi\to\gamma f_0(1710)\to\gamma \omega\omega$ & $0.31\pm0.06\pm0.08$ \cite{BES:omegaomega} & $$ \\
$J/\psi\to\gamma f_0(1710)\to\gamma \eta\eta$ & $2.35^{+0.13+1.24}_{-0.11-0.74}$ \cite{BES:etaeta} & $$ \\
\end{tabular}
\end{ruledtabular}
\end{table}

In Model I, one may argue that the constructive interference between the $s\bar s$ and glueball components can lead to a large radiative $J/\psi$ rate for $f_0(1710)$. On the other hand, since $|f_0(1500)\ra=-0.41|N\ra+0.35|S\ra-0.84|G\ra$ in this model, it is clear that the radiative $J/\psi$ decay to $f_0(1500)$ is mainly governed by its glueball content as the constructive and destructive interferences between the $q\bar q$ and glueball components tend to cancel each other. Therefore, it will be difficult to understand why $J/\psi \to \gamma f_0(1500)$ is largely suppressed relative to $f_0(1710)$ if $f_0(1500)$ is primarily a glueball.

\subsection{Ratio of $f_0$ decays to $\pi\pi$, $K\ov K$ and $\eta\eta$}

Since glueballs are flavor singlets, their decays are naively expected to be flavor symmetric. For example, considering a pure glueball decay into $\pi\pi$ and $K\ov K$, we have
\be \label{eq:RG}
R(G)\equiv {\Gamma(G\to \pi\pi)\over\Gamma(G\to K\!\bar K)}={3\over
 4}\left({g^{\pi\pi}\over g^{K\!\bar{K}}}\right)^2{p_\pi\over p_K},
\en
where the glueball couplings to two pseudoscalar mesons are expected to be flavor independent, namely, $g^{K\!\bar K}=g^{\pi\pi}$. In the SU(3) limit, $R(G)=3/4$. Taking into account of phase space corrections, we find $R(G)=0.90$ and 0.98 for $M_G=1710$ MeV and 1500 MeV, respectively.

However, the above argument is no longer true for scalar glueballs due to chiral suppression. It was noticed long time ago by Carlson {\it et al.} \cite{Carlson}, by Cornwall and Soni \cite{Cornwall} and revitalized recently by Chanowitz \cite{Chanowitz} that a scalar glueball cannot decay into a quark-antiquark pair in the chiral limit, i.e., $A(G\to q\bar q)\propto m_q$. Consequently, scalar glueballs should have larger coupling to $K\ov K$ than to $\pi\pi$. Nevertheless, chiral suppression for the ratio $\Gamma(G\to \pi\pi)/\Gamma(G\to K\bar K)$ at the hadron level should not be so strong as the current quark mass ratio $m_u/m_s$.
It has been suggested \cite{Chao} that $m_q$ should be interpreted as the scale of chiral symmetry breaking.
A precise estimate of the chiral suppression effect is a difficult issue because of the hadronization process from $G\to q\bar q$ to $G\to\pi\pi$ and the possible competing $G\to q\bar qq\bar q$ mechanism is not well-known \cite{Carlson,Chao,Jin,Chanowitz:reply}. The only reliable method for tackling with the nonperturbative effects is lattice QCD. An earlier lattice calculation \cite{Sexton} did support the chiral-suppression effect with the result
\be \label{eq:chiralsupp}
g^{\pi\pi}:g^{K\!\bar K}:g^{\eta\eta}=0.834^{+0.603}_{-0.579}:2.654^{+0.372}_{-0.402}:
3.099^{+0.364}_{-0.423},
\en
which are in sharp contrast to the flavor-symmetry limit with
$g^{\pi\pi}:g^{K\bar K}:g^{\eta\eta}=1:1:1$.
Although the errors are large, the lattice result did show a sizable deviation from the flavor-symmetry limit. Therefore,
$\Gamma(G\to \eta\eta)>\Gamma(G\to K\ov K)\gg \Gamma(G\to\pi\pi)$.

The experimental results
\be
R(f_0(1710))\equiv {\Gamma(f_0(1710)\to\pi\pi)\over \Gamma(f_0(1710)\to K\ov K)}=\cases{ <0.11 & BESII from $J/\psi\to\omega(K\ov K,\pi\pi)$ \cite{BES:omegaKK}, \cr
                 0.20\pm 0.04 & WA102 \cite{WA102}, \cr
                 0.31\pm0.05 & CLEO  \cite{Dobbs:2015dwa}, \cr
                 0.32\pm0.14 & Albaladejo and Oller \cite{Albaladejo:2008qa}, \cr
                 0.41^{+0.11}_{-0.17} & BESII from $J/\psi\to\gamma(K\ov K,\pi\pi)$ \cite{BES:R}. \cr}
\en
clearly indicate that the $\pi\pi$ production in $f_0(1710)$ decays is largely suppressed relative to $K\ov K$. Theoretically, the ratio of $\pi\pi$ and $K\ov K$ productions in $f_{0i}$ decays is given by \cite{CCL}
\be \label{eq:f0topipiKK}
 R(f_{0i})\equiv {\Gamma(f_{0i}\to \pi\pi)\over\Gamma(f_{0i}\to K\!\ov K)}=
 3\left({\alpha_i/\sqrt{2}+g^{\pi\pi}\gamma_i\over r_a\alpha_i/\sqrt{2}+\beta_i+2g^{K\!\bar{K}}\gamma_i}\right)^2\,{p_\pi\over p_K},
\en
where $\alpha_i,\beta_i$ and $\gamma_i$ are the coefficients of the $f_{0i}$ wave function defined in Eq. (\ref{eq:f0}), $p_h$ is the c.m. momentum of the hadron $h$ and the parameter $r_a$ denotes a possible SU(3) breaking effect in the
OZI allowed decays when the $s\bar s$ pair is created relative to
the $u\bar{u}$ and $d\bar{d}$ pairs.
In Model II,  $f_0(1710)$ has the smallest content of $s\bar s$ (see Eq. (\ref{eq:CCL})) even though it decays dominantly to $K\ov K$; the smallness of $R(f_0(1710))$ arises from the chiral suppression of scalar glueball decay.
Specifically, the parameters $g^{\pi\pi}=0.12$, $g^{K\!\bar K}=3.15\, g^{\pi\pi}$ and $r_a=1.22$ were chosen in \cite{CCL}. The ratio $g^{\pi\pi}:g^{K\!\bar K}=1:3.15$ is consistent with the lattice calculation (\ref{eq:chiralsupp}). Substituting Eq. (\ref{eq:CCL}) into Eq. (\ref{eq:f0topipiKK}) leads to $R(f_0(1710))=0.31^{+0.11}_{-0.03}$\,.

Note that in the absence of chiral suppression the smallness of $R(f_0(1710))$ can be naturally explained in terms of the large $s\bar s$ component of $f_0(1710)$ in Model I. For example, we found $R(f_0(1710))=0.22$ for $r_a=1$ and $g^{\pi\pi}=g^{K\!\bar K}=1$. However, the presence of chiral suppression will render the ratio even smaller. If we apply the same parameters  $g^{\pi\pi}=0.12$, $g^{K\!\bar K}=3.15\, g^{\pi\pi}$ and $r_a=1.22$ as in Model II, we will obtain $R(f_0(1710))=0.025$ which is too small compared to experiment. Hence, if the chiral suppression effect is confirmed in the future, this will favor Model II over Model I.

Although $f_0(1500)$ in Model II has the largest content of $s\bar s$, the $K\ov K$ production is largely suppressed relative to $\pi\pi$ due to the destructive interference between $n\bar n$ and $s\bar s$ components
\be
R(f_0(1500))\approx 3\left({\alpha_2\over r_a\alpha_2+\sqrt{2}\beta_2}\right)^2 {p_\pi\over p_K}=3.9\left({\alpha_2\over r_a\alpha_2+\sqrt{2}\beta_2}\right)^2.
\en
The experimental value of $4.1\pm0.5$ for $R(f_0(1500))$ \cite{PDG} can be fitted with two possible solutions
\be
{\alpha_2\over r_a\alpha_2+\sqrt{2}\beta_2}\approx \pm 1\,.
\en
Setting $r_a=1$ for the moment, we are led to $\beta_2\approx 0$ or $\beta_2/\alpha_2\approx -\sqrt{2}$\,. The second solution is nothing but a flavor octet $f_0(1500)$ as advocated in Model II before. With a small SU(3) breaking in the parameter $r_a$, namely, $r_a=1.22$, we obtain $R(f_0(1500))\approx 4.1$ in excellent agreement with experiment.
\footnote{After taking into account the contribution from the glueball content, we obtain $R(f_0(1500))=3.7\pm0.6$.}
The above discussion explains why the measurement of $R(f_0(1500)$ favors the flavor octet nature of $f_0(1500)$.

In Model I, $f_0(1500)$ is dominated by the glueball content. Since $R(G)$ is of order unity for flavor-independent couplings, one needs a large $q\bar q$ mixing with the glueball component in order to accommodate the experimental result of $R(f_0(1500))$ in this model. The destructive interference between the $n\bar n$ and $s\bar s$ components have to be adjusted in such a way that the production of the $K\!\ov K$ pair is severely suppressed so that the quark component alone  will lead to a very huge $R(f_0(1500))$ to compensate for the smallness of $R(f_0(1500))$ produced by the glueball component. From Eq. (\ref{eq:f0topipiKK}) with $r_a=1$ and $g^{\pi\pi}=g^{K\!\bar K}=1$ and the wave function  $|f_0(1500)\ra=-0.41|N\ra+0.35|S\ra-0.84|G\ra$, we find
$R(f_0(1500))=1.9$ which is slightly smaller than the value of 2.4 obtained in \cite{Close2}. At any rate, the predicted ratio $R(f_0(1500))$ is still smaller than experiment.

Can the experimental ratio $R(f_0(1500))$ be accommodated in Model I ? To see this, we notice that
\be \label{}
 R(f_0(1500))\approx
 3.9\left({r_a \alpha_2/\sqrt{2}+g^{\pi\pi}\gamma_2\over r_a\alpha_2/\sqrt{2}+\beta_2+2g^{K\!\bar K}\gamma_2}\right)^2.
\en
Taking $r_a=1$ and $g^{\pi\pi}=g^{K\!\bar K}=1$,  the experimental measurement can be accommodated by having either  $\beta_2+\gamma_2\approx 0$ or  $\sqrt{2}\alpha_2+\beta_2+3\gamma_2\approx 0$.
Neither of the relations can be satisfied in Model I with $\alpha_2=-0.41$, $\beta_2=0.35$ and $\gamma_2=-0.84$\,. In principle, one can introduce chiral suppression to accommodate $R(f_0(1500))$. For example, $g^{\pi\pi}=0.0623$, $g^{K\!\bar K}=3.15\, g^{\pi\pi}$ and $r_a=1.22$ will lead to $R(f_0(1500))=4.1$\,. However, the same set of parameters also leads to a too small ratio $R(f_0(1710))=0.020$\,. In other words, it is difficult to explain the ratios of $\pi\pi$ and $K\ov K$ productions in $f_0(1500)$ and $f_0(1710)$ decays simultaneously in Model I.

We next turn to the $\eta\eta$ modes and consider two ratios that have been measured: $R^{f_0(1710)}_{\eta\eta/K\!\bar K}\equiv\Gamma(f_0(1710)\to \eta\eta)/\Gamma(f_0(1710)\to K\!\ov K)$ and $R^{f_0(1500)}_{\eta\eta/\pi\pi }\equiv\Gamma(f_0(1500)\to \eta\eta)/\Gamma(f_0(1500)\to \pi\pi)$. Their theoretical expressions are given by \cite{CCL}
\be \label{eq:Retaeta}
 R^{f_0(1710)}_{\eta\eta/K\!\bar K} &=&
 \left({a_\eta^2\alpha_3/\sqrt{2}+r_a b_\eta^2\beta_3+g^{\eta\eta}(a_\eta^2+b_\eta^2)\gamma_3+\rho_{ss}(2a_\eta^2+b_\eta^2+{4\over\sqrt{2}}a_\eta b_\eta)\gamma_3\over r_a\alpha_3/\sqrt{2}+\beta_3+2g^{K\!\bar{K}}\gamma_3}\right)^2\,{p_\eta\over p_K},  \non \\
 R^{f_0(1500)}_{\eta\eta/\pi\pi } &=&{1\over 3}
 \left({a_\eta^2\alpha_2/\sqrt{2}+r_a b_\eta^2\beta_2+g^{\eta\eta}(a_\eta^2+b_\eta^2)\gamma_2+\rho_{ss}(2a_\eta^2+b_\eta^2+{4\over\sqrt{2}}a_\eta b_\eta)\gamma_2\over r_a\alpha_2/\sqrt{2}+g^{\pi\pi}\gamma_2}\right)^2\,{p_\eta\over p_\pi},
\en
where
\be
a_\eta={\cos\theta-\sqrt{2}\sin\theta\over \sqrt{3}}, \qquad b_\eta=-{\sin\theta+\sqrt{2}\cos\theta\over \sqrt{3}},
\en
with $\theta$ being the $\eta-\eta'$ mixing angle defined by
\be
\eta=\eta_8\cos\theta-\eta_0\sin\theta,\qquad
\eta'=\eta_8\sin\theta+\eta_0\cos\theta.
\en
In Eq. (\ref{eq:Retaeta}), the coupling $\rho_{ss}$ is the ratio of the doubly OZI suppressed coupling to that of the OZI allowed one \cite{CCL}.

\begin{table}[t]
\caption{The ratios of $f_0(1710)$ to $\eta\eta$ and $K\!\ov K$ and $f_0(1500)$ to $\eta\eta$ and $\pi\pi$. As stated in the text, the PDG value of $0.145\pm0.027$ for ${\Gamma(f_0(1500)\to\eta\eta) \over \Gamma(f_0(1500)\to \pi\pi)}$
comes from the fit to three different experimental measurements.} \label{tab:etaeta}
\begin{tabular}{ | l c c c |}  \hline
 ~~~ & ~~~~Expt~~~~ & ~~~~Model I~~~~~ & ~~~~Model II~~~~~\\ \hline
 ~~~${\Gamma(f_0(1710)\to\eta\eta) \over \Gamma(f_0(1710)\to K\!\bar K)}$ & $0.48\pm0.15$ \cite{WA102} & 0.24 & $0.52^{+0.33}_{-0.34}$ \\
 ~~~${\Gamma(f_0(1500)\to\eta\eta) \over \Gamma(f_0(1500)\to \pi\pi)}$ & $0.145\pm0.027$ \cite{PDG} & 0.19 & $0.078^{+0.025}_{-0.027}$ \\ \hline
\end{tabular}
\end{table}

Using the mixing angle $\theta=-14.4^\circ$, $g^{\eta\eta}=4.74 g^{\pi\pi}$ \cite{CCL} and $\rho_{ss}=0.60^{+1.24}_{-2.02}$, the predicted ratios in Models I and II are exhibited in Table \ref{tab:etaeta}.
We see that Model II gives a better description of $R^{f_0(1710)}_{\eta\eta/K\!\bar K}$, while Model I seems to yield a better agreement for $R^{f_0(1500)}_{\eta\eta/\pi\pi}$. Note that the PDG value of $0.145\pm0.027$ \cite{PDG} for the latter ratio
comes from the fit to the three measurements ranging from $
0.230\pm0.097$ \cite{Amsler95} to $0.18\pm0.03$ \cite{Barberis00}
and $0.080\pm0.033$ \cite{AmslerCB}. As a result, the prediction of Model II is consistent with one of the experiments. Therefore, it is important to have
an improved measurement of $R^{f_0(1500)}_{\eta\eta/\pi\pi}$ in the future.

Finally, we would like to remark that in our mixing model we rely on the measurements of two-body decays of $f_0(1500)$ and $f_0(1710)$. There is no use of the branching fractions of $f_0(1370)$. As explained in \cite{CCL}, the measurements of $\Gamma(f_0(1370)\to\pi\pi)/\Gamma(f_0(1370)\to K\!\ov K)$ and $\Gamma(f_0(1370)\to\eta\eta)$, for example, span a large range from different experiments. Therefore, they are not employed as the fitting input. Nevertheless, the $f_0(1370)$ mass is used for a best $\chi^2$ fit. We also use its mass to fix the parameter $x$ in our model. Since there are three quarkonium states $|U\ra$, $|D\ra$, $|S\ra$ and one pure glueball state $|G\ra$, it is necessary to include $f_0(1370)$ to form the scalar meson basis in addition to $a_0(1450)$, $f_0(1500)$ and $f_0(1710)$.

\subsection{Ratio of $J/\psi$ decays to $f_0(1710)\omega$ and $f_0(1710)\phi$}
The ratio of $J/\psi$ decays to $f_0(1710)\omega$ and $f_0(1710)\phi$ provides another useful test on the mixing-matrix models. Experimentally,
\be
{\Gamma(J/\psi\to \omega f_0(1710))\over \Gamma(J/\psi\to \phi f_0(1710))}={\Gamma(J/\psi\to \omega f_0(1710)\to\omega K\!\bar K)\over \Gamma(J/\psi\to \phi f_0(1710)\to \phi K\!\bar K)}=\cases{ 3.3\pm1.3 & BES \cite{BES:omegaKK},
\cr 1.3\pm0.4 & DM2 \cite{Falvard}. \cr}
\en
Hence, $J/\psi\to\omega
f_0(1710)$ tends to have a rate larger than $J/\psi\to \phi f_0(1710)$. This
is easily understood in Model II because the $n\bar n$ content is more copious than $s\bar s$ in $f_0(1710)$. Indeed, the prediction of $\Gamma(J/\psi\to
\omega f_0(1710))/\Gamma(J/\psi\to \phi f_0(1710))=4.1$ \cite{CCL} is
consistent with the BES measurement. If
$f_0(1710)$ is dominated by $s\bar s$ as advocated in Model I, one will naively expect a suppression of the
$\omega f_0(1710)$ production relative to $\phi f_0(1701)$. One
way to circumvent this apparent contradiction with experiment is
to assume a large OZI violating effect in the scalar meson
production~\cite{Close2}. That is, the doubly OZI suppressed
process (i.e. doubly disconnected diagram) is assumed to dominate
over the singly OZI suppressed (singly disconnected) process
\cite{Close2}. In contrast, a larger $\Gamma(J/\psi\to \omega
f_0(1710))$ rate over that of $\Gamma(J/\psi\to \phi f_0(1710))$
is naturally accommodated in Model II without asserting large
OZI violating effects.

\subsection{Scalar resonance contributions to $B_s\to J/\psi\pi^+\pi^-$}
Resonant structure of $B_s\to J/\psi\pi^+\pi^-$ has been studied recently by Belle \cite{Belle:Bs} and LHCb \cite{LHCb:Bs,LHCb:Bs2}. For the scalar resonances, Belle made the first observation of $B_s\to J/\psi f_0(980)$ and the first evidence for $B_s\to J/\psi f_0(1370)$ with $M=1405\pm15^{+1}_{-7}$ MeV and $\Gamma=54\pm33^{+14}_{-13}$ MeV. The resonance state with mass $1475.1\pm6.3$ MeV and width $112.7\pm11.1$ MeV observed by LHCb  was originally identified with $f_0(1370)$ in the LHCb analysis \cite{LHCb:Bs}, but it was then assigned to $f_0(1500)$ in the latest LHCb study  \cite{LHCb:Bs2}. The possible resonances considered by LHCb include $f_0(500), f_0(980), f_2(1270), f_0(1500), f'_2(1525), f_0(1710), f_0(1790)$ and $\rho(770)$.
LHCb has carried out two different fits for the fit fractions of various scalar resonances. In Table \ref{tab:LHCb} we list the fit fractions for $f_0(980),f_0(1500)$ and $f_0(1790)$.

Because of the spectator $s$ quark of $B_s$, the isosinglet scalar resonance $f_0$ produced in $B_s\to J/\psi f_0$ decays should have a sizable $s\bar s$ component. It is well known that $f_0(980)$ is dominated by $s\bar s$.
Indeed, we learn from Table \ref{tab:LHCb} that $B_s\to J/\psi f_0(980)$ has the largest rate among all the scalar resonances under consideration. Moreover, the $s\bar s$ component of $f_0(1500)$ should be more abundant than that of $f_0(1710)$.

\begin{table}[h]
\caption{Fit fractions (\%) of contributing scalar resonancs to $B_s\to J/\psi \pi^+\pi^-$ for solutions I and II \cite{LHCb:Bs2}. Only the dominant states $f_0(980),f_0(1500)$ and $f_0(1790)$ are shown here. Non-resonant contributions exist in Solution II but not in Solution I.} \label{tab:LHCb}
\begin{tabular}{ | l c c  |}  \hline
 ~~~Component & ~~~~Solution I~~~~ & ~~~~Solution II~~~~~ \\ \hline
 ~~~$f_0(980)$ & $70.3\pm1.5^{+0.4}_{-5.1}$ & $92.4\pm2.0^{+~0.8}_{-16.0}$ \\
 ~~~$f_0(1500)$ & $10.1\pm0.8^{+1.1}_{-0.3}$ & $9.1\pm0.9\pm0.3$  \\
 ~~~$f_0(1790)$ & $2.4\pm0.4^{+5.0}_{-0.2}$ & $0.9\pm0.3^{+2.5}_{-0.1}$  \\ \hline
\end{tabular}
\end{table}

It is expected in Model I that the production of $f_0(1710)$ in $B_s\to J/\psi f_{0i}\to J/\psi\pi^+\pi^-$ decays to be more prominent than $f_0(1500)$ and $f_0(1370)$ and the other way around in Model II. To quantity this statement, we note that
the relative production rates of $f_{0i}$ are
\be
\Gamma(B_s\to J/\psi f_0(1370):f_0(1500):f_0(1710)) = 
\cases{1:25: 177 & Model I,  \cr
1: 2.7: 0.12 & Model II. \cr}
\en
Using the narrow width approximation,
\footnote{It is known that the narrow width approximation works provided that the resonance is not too broad. To check the validity of Eq. (\ref{eq:NWA}), we can define a quantity $\eta$
\be
\eta={\Gamma(B_s\to J/\psi f_0\to J/\psi\pi\pi)\over \Gamma(B_s\to J/\psi f_0){\cal B}(f_0\to \pi\pi)} . \nonumber
\en
The deviation of $\eta$ from unity will give a measure of the violation of the narrow width approximation. Assuming $q^2$ independence of the weak matrix element $\langle J/\psi f_0|H_W|B_s\rangle$ and the strong coupling $g_{f_0\pi\pi}$ and using the formula given in \cite{Cheng:2002mk}, we find that $\eta$ is indeed close to unity, $\eta=0.95$ for $\Gamma(f_0(1500))=109$ MeV and $\eta=0.93$ for $\Gamma(f_0(1710))=135$ MeV.}
\be \label{eq:NWA}
\Gamma(B_s\to J/\psi f_0\to J/\psi \pi\pi)=\Gamma(B_s\to J/\psi f_0)\B(f_0\to \pi\pi),
\en
and the branching fractions $\B(f_0(1500)\to\pi\pi)=0.349\pm0.023$ \cite{PDG}, $\B(f_0(1710)\to K\ov K)=0.36\pm0.12$ and $R(f_0(1710))=0.32\pm0.14$ \cite{Albaladejo:2008qa}, we obtain
\be \label{eq:ratio}
&& \Gamma(B_s\to J/\psi f_0(1500)\to J/\psi\pi\pi): \Gamma(B_s\to J/\psi f_0(1710)\to J/\psi\pi\pi)\non \\
&& =\Gamma(B_s\to J/\psi f_0(1500))\B(f_0(1500)\to\pi\pi):\Gamma(B_s\to J/\psi f_0(1710))\B(f_0(1710)\to\pi\pi) \non \\
&& \approx\cases{ 1: 2.33 & Model I, \cr
1: 0.015 & Model II. \cr}
\en
Due to the unknown branching fraction of  $f_0(1370)\to\pi\pi$, we have not included $f_0(1370)$ in the above equation and for simplicity we have only considered the central values of $\B(f_0(1500)\to\pi\pi)$ and $\B(f_0(1710)\to\pi\pi)$ and ignored phase-space corrections. Moreover, we have not taken account the contributions to $B_s\to J/\psi f_{0i}$ from the glueball component of $f_{0i}$ through glueball-$s\bar s$ mixing. However, it will not modify the pattern shown in Eq. (\ref{eq:ratio}).
Evidently, Model II is preferred by the data while Model I is not favored because the measured $\pi^+\pi^-$ spectrum is peaked near the invariant mass $M(\pi^+\pi^-)=1.50$ GeV and its rate is much higher than that at $M(\pi^+\pi^-)=1.71$ GeV (see Figs. 16 and 17 of \cite{LHCb:Bs2}). Hence, we conclude that the LHCb data on the scalar resonance contributions to $B_s\to J/\psi\pi^+\pi^-$ imply the $s\bar s$ content abundant in $f_0(1500)$ and negligibly small in $f_0(1710)$.

By the same token, it is expected that the scalar contributions to $B_s\to J/\psi K^+K^-$ lead to the following pattern
\be \label{}
&& \Gamma(B_s\to J/\psi f_0(1500)\to J/\psi K\ov K): \Gamma(B_s\to J/\psi f_0(1710)\to J/\psi K\ov K)\non \\
&& \approx\cases{ 1: 29.6 & Model I, \cr
1: 0.19 & Model II, \cr}
\en
where use of $\B(f_0(1500)\to K\ov K)=0.086\pm0.010$ \cite{PDG} has been made. This can be studied by LHCb in the near future to test Models I and II.

\subsection{Near mass degeneracy of $a_0(1450)$ and $K_0^*(1430)$}
SU(3) symmetry leads naturally to the near
degeneracy of $a_0(1450)$, $K_0^*(1430)$ and $f_0(1500)$. However,
in order to accommodate the observed branching ratios of strong
decays, SU(3) symmetry needs to be broken slightly in the
mass matrix and/or in the decay amplitudes. One also needs
$M_S>M_U=M_D$ a little bit in order to lift the degeneracy of $a_0(1450)$ and
$f_0(1500)$.

In Model I, $M_S-M_N=317\pm25$ MeV in fit 1 and $378\pm8$ MeV in fit 2 \cite{Close2}. Therefore, it cannot explain the near mass degeneracy in this model. In Model II, $M_S-M_N=25$ MeV which is much smaller than the constituent quark masses.

\subsection{$f_0$ production in $\gamma\gamma$ reaction}
The scalar meson $f_0(1500)$ was not seen in $\gamma\gamma\to K_SK_S$ by L3 \cite{L3}, nor in $\gamma\gamma\to\pi^+\pi^-$ by ALEPH \cite{Aleph}. However, a resonance
observed in $\gamma\gamma\to\pi^0\pi^0$ by Belle \cite{Uehara:2008} is close to the $f_0(1500)$ mass, though it is also consistent with $f_0(1370)$ because of the large errors in the experiment and the large uncertainty in the $f_0(1370)$ mass. $f_0(1710)$ has been seen in $\gamma\gamma\to K_SK_S$ \cite{Uehara:2013,L3}. The $2\gamma$ couplings are sensitive to the glueball mixing with $q\bar q$. In general, we have
\be \label{eq:2gamma}
\Gamma(f_{0i}\to\gamma\gamma)\propto \left(\alpha_{i}{5\over 9\sqrt{2}}+\beta_{i}{1\over 9}\right)^2.
\en
It follows that
\be \label{eq:2gammarate}
\Gamma_{f_0(1370)\to\gamma\gamma}: \Gamma_{f_0(1500)\to\gamma\gamma}: \Gamma_{f_0(1710)\to\gamma\gamma} = \cases{ 8.9 : 1.0 : 1.6  & Model I, \cr
                                           9.3: 1.0 : 1.7 & Model II, \cr}
\en
apart from phase space factors.
Hence, the absence of $f_0(1500)$ in $\gamma\gamma$ reactions does not necessarily imply a glueball content for $f_0(1500)$. Note that in Model II $f_0(1500)$ has the smallest $2\gamma$ coupling of the three states even though it has the least glue content.
Indeed, it is known that the weak $2\gamma$ coupling is not a good criterion to test the nature of a glueball because the $q\bar q$ state can also have a weak coupling to two photons by adjusting the coefficients $\alpha_i$ and $\beta_i$.

\subsection{$f_0$ production in $p\bar p$ collision}
Crystal Barrel did not see $f_0(1710)$ in $p\bar p\to \eta\eta\pi^0$ \cite{AmslerCB}. This non-observation of $f_0(1710)$ in $p\bar p$ has been used to argue that it is $s\bar s$ dominated. However, this argument is moot since the analysis of \cite{Anisovich} based on WA102 data and Fermilab E835 experiment \cite{E835} saw both $f_0(1500)$ and $f_0(1710)$  in $p\bar p\to \eta\eta\pi^0$ .

\section{Nearby resonances}
In the vicinity of $f_0(1710)$ there exist several other $0^{++}$ states such as $f_0(1790)$ and $X(1812)$, $f_0(2020)$ and $f_0(2100)$. The former was seen in $J/\psi\to \phi\pi^+\pi^-$ by BESII with mass $1790^{+40}_{-30}$ MeV and width $270^{+60}_{-30}$ MeV \cite{f01790}. The $X(1812)$ state  was seen in the doubly OZI-suppressed decay $J/\psi\to\gamma\omega\phi$ by BESII with mass $1812^{+19}_{-26}\pm18$ MeV and width $105^{+20}_{-28}$ MeV \cite{X1812} and confirmed by BESIII with mass $1795\pm7^{+13}_{-~5}\pm19$ MeV and width $95\pm10^{+21}_{-34}\pm75$ MeV \cite{X1810}. Although the large width of $f_0(1790)$ has a strong overlap with $f_0(1710)$, there is a clear distinction between the two resonances: $f_0(1790)$ is reconstructed mainly in pion decay modes and couples weekly to $K\ov K$, whereas $f_0(1710)$ is reconstructed predominantly in kaon decay channels. However, the existence of the former has never been confirmed by other experiments.


If $f_0(1790)$ and $X(1812)$ are supposed to be truly new states distinct from $f_0(1710)$, then the question is how to accommodate these two new states out of $n\bar n$, $s\bar s$ and $G$? The addition of these two states
into the picture requires an enlargement of the basis. In
QCD, the next simplest states having the quantum numbers
compared with the quarkonia and glueball basis are the
hybrid basis composed of an antiquark $\bar q$, a quark $q$, and a
gluon $g$, i.e. $q\bar qg$ which contains two independent $0^{++}$
states, $(u\bar u + d\bar d)g/\sqrt{2}$
and $s\bar sg$. It has been proposed in \cite{He:2006} that they are scalar hybrids: $f_0(1790)$ is primarily $(u\bar u + d\bar d)g/\sqrt{2}$, while $X(1812)$ is a $s\bar sg$ hybrid state. The analysis of \cite{He:2006} seems to imply that the mixing pattern, for example, Eq. (\ref{eq:Close}), is not affected by the extra new states.


\newlength{\myheight}
\setlength{\myheight}{1cm}
\newlength{\myheighta}
\setlength{\myheighta}{2cm}
\newlength{\myheightb}
\setlength{\myheightb}{0.8cm}

\begin{table}[t]
\begin{ruledtabular}
\centering \caption{Comparison of two different types of models for the mixing matrices of the isosinglet scalar mesons $f_0(1370)$, $f_0(1500)$ and $f_0(1710)$.
Experimental results are taken from Sec. III.} \label{tab:summary}
\begin{tabular}{l | c |c }
 Experiment & Model I \cite{Close2} & Model II \cite{CCL} \\
 \hline
 \parbox[c][\myheighta][c]{0cm}{}
 $\left(\matrix{ |f_0(1370)\ra \cr |f_0(1500)\ra \cr |f_0(1710)\ra \cr}\right)=(...)\left(\matrix{|N\ra \cr
 |S\ra \cr |G\ra \cr}\right)$ & $\left( \matrix{ -0.91 & -0.07 & 0.40 \cr
                 -0.41 & 0.35 & -0.84 \cr
                0.09 & 0.93 & 0.36 \cr
                  } \right)$ & $\left( \matrix{ 0.78(2) & 0.52(3) & -0.36(1) \cr
                 -0.55(3) & 0.84(2) & 0.03(2) \cr
                0.31(1) & 0.17(1) & 0.934(4) \cr
                  } \right)$  \\
\hline
 Mass of the lightest scalar $G$ & $M_G\sim 1464-1519 $ MeV & $M_G \sim 1665$ MeV \\
 in LQCD $\sim {\cal O}(1700)$MeV & & \\
\hline
   \parbox[c][\myheight][c]{0cm}{}
 ${\Gamma(J/\psi\to f_0(1710)\gamma)\over \Gamma(J/\psi\to f_0(1500)\gamma)}\sim {\cal O}(10)$ & If $f_0(1500)$ is primarily a glueball,  & Yes, as $|f_0(1710)\ra\sim |G\ra$   \\
 & this ratio will be less than 1. & \\
 \hline
 \parbox[c][\myheight][c]{0cm}{}
 ${\Gamma(f_0(1710)\to \pi\pi)\over\Gamma(f_0(1710)\to K\!\bar K)}=0.31\pm0.05$ & $f_0(1710)$ dominated by $s\bar s$  & Chiral suppression \\
 \hline
  \parbox[c][\myheightb][c]{0cm}{}
  & If $f_0(1500)$ is primarily a glueball, & Well explained with the \\
 ${\Gamma(f_0(1500)\to \pi\pi)\over\Gamma(f_0(1500)\to K\!\bar K)}=4.1\pm0.5$ &  this ratio will be of order unity.  & flavor octet structure of  \\
 & Needs a large mixing with $q\bar q$. & $f_0(1500)$. \\
 \hline
  \parbox[c][\myheight][c]{0cm}{}
 ${\Gamma(f_0(1710)\to \eta\eta)\over\Gamma(f_0(1710)\to K\!\bar K)}=0.48\pm0.15$ &  0.24  & $0.52^{+0.33}_{-0.34}$  \\
 \hline
  \parbox[c][\myheighta][c]{0cm}{}
 ${\Gamma(f_0(1500)\to \eta\eta)\over\Gamma(f_0(1500)\to \pi\pi)}=\cases{0.230\pm0.097 \cr 0.18\pm0.03 \cr 0.080\pm0.033 \cr}$ &  0.19  & $0.078^{+0.025}_{-0.027}$  \\
 \hline
  \parbox[c][\myheight][c]{0cm}{}
 ${\Gamma(J/\psi\to f_0(1710)\omega)\over \Gamma(J/\psi\to f_0(1710)\phi)}=\cases{3.3\pm1.3 \cr 1.3\pm0.4\cr}$ & The ratio is naively less than 1. & Yes, as $|S\ra$ is small \\
 & Needs large OZI-violating effects.  & in $f_0(1710)$  \\
 \hline
 Non-observation of $f_0(1710)$   & Dominant $f_0(1710)$ production  & Dominant $f_0(1500)$ production,  \\
 and observation of $f_0(1500)$  & followed by $f_0(1500)$   & while $f_0(1710)$ is negligible \\
 in $B_s\to J/\psi\pi^+\pi^-$ by LHCb & &  \\
 \hline
  Near mass degeneracy of  & No, it cannot be explained & Yes, as $M_S-M_N\approx$ 25 MeV \\
  $a_0(1450)$ and $K_0^*(1430)$ & as $M_S-M_N\approx$ 200-300 MeV & \\
 \hline
 $f_0(1500)$ not seen in $\gamma\gamma$  &  & \\
 reactions except probably  & See Eq. (\ref{eq:2gammarate}) & See Eq. (\ref{eq:2gammarate})  \\
 in $\gamma\gamma\to\pi^0\pi^0$ & &  \\
 \end{tabular}
\end{ruledtabular}
 \end{table}

\section{Discussion and Conclusions}

In this work we have considered lattice calculations and experimental data to infer the glue and $q\bar q$ components of the isosinglet scalar mesons. The scalar glueball mass calculated in quenched and unquenched lattice QCD and the experimental measurement the radiative decay $J/\psi\to\gamma f_0$ clearly indicate a dominant glueball component in $f_0(1710)$. The measured ratio of $f_0(1710)$ decays to $\pi\pi$ and $K\ov K$ implies the importance of chiral suppression effects in scalar glueball decays to two pseudoscalar mesons.
The LHCb data on the scalar resonance contributions to $B_s\to J/\psi\pi^+\pi^-$ imply the $s\bar s$ content abundant in $f_0(1500)$ and negligible in $f_0(1710)$. The observed ratio of $J/\psi$ decays to $f_0(1710)\omega$ and $f_0(1710)\phi$ suggests that the $n\bar n$ component of $f_0(1710)$ should be more copious than the $s\bar s$ one.   The near mass degeneracy of $a_0(1450)$ and $K_0^*(1430)$ demands a small mass difference between the model parameters $M_S$ and $M_N$. We have shown explicitly that if $f_0(1500)$ is dominated by the $q\bar q$ components, then the experimental ratio of $f_0(1500)$ decays to $\pi\pi$ and $K\ov K$ will require $f_0(1500)$ be predominately a flavor octet. This is consistent with the near degeneracy of $a_0(1450)$ and $K_0^* (1430)$.
The comparison of two different types of models for the mixing matrices of the isosinglet scalar mesons is summarized in Table \ref{tab:summary}.

It was originally argued that $f_0(1500)$ is primarily a glueball because the $q\bar q$ state cannot explain the ratio $R(f_0(1500))$ and its weak production in $\gamma\gamma$ reactions simultaneously. However, this argument is no longer valid in Model II where $f_0(1500)$ is predominantly a flavor octet $q\bar q$ state.  The ratio $R(f_0(1500))$ and its weak coupling with two photons are well explained. We have pointed out that in Model I it is difficult to
explain the ratios of $\pi\pi$ and $K\ov K$ productions in $f_0(1500)$ and $f_0(1710)$ decays simultaneously. In principle, one can introduce chiral suppression to accommodate the measured $R(f_0(1500))$, but the same effect will also lead to a too small $R(f_0(1710))$.
Moreover, Model I cannot naturally explain the ratio of the radiative $J/\psi$ decays to $f_0(1710)$ and $f_0(1500)$, the ratio of $J/\psi$ decays to $f_0(1710)\omega$ and $f_0(1710)\phi$, and the sizable $f_0(1500)$ contributions to $B_s\to J/\psi\pi^+\pi^-$.

Chiral suppression plays an essential role in distinguishing the glueball from the $q\bar q$ components. Because of the chiral suppression effect for the scalar glueball decays, $R(G)$ is naturally small. The observation of $R(f_0(1710))\ll 1$ and $R(f_0(1500))\gg 1$ clearly suggests that $f_0(1700)$ is most likely to have a large glue component, whereas $f_0(1500)$ is dominated by the quark content.

We conclude that all the analyses in this work suggest the prominent glueball nature of $f_0(1710)$ and the flavor octet structure of $f_0(1500)$.

\section{Acknowledgments}

We would like to thank Jean-Marie Fr\`ere for useful comments on pseudoscalar glueballs.
This research was supported in part by the Ministry of Science and Technology of R.O.C. under Grant
Nos. 103-2112-M-001-005 and  103-2112-M-033-002-MY3 and  and the USDOE grant
DE-FG05-84ER40154.


\begin{thebibliography}{99}
\bibitem{Frere} We wish to thank Jean-Marie Fr\`ere for pointing out this to us.
\end{thebibliography}
\end{document}